# Assessing Privacy Leakage in Synthetic 3-D PET Imaging using Transversal GAN

Robert V. Bergen[a*], Jean-Francois Rajotte[a], Fereshteh Yousefirizi[b], Arman Rahmim[b,c], Raymond T. Ng[a]

[a]Data Science Institute, University of British Columbia, BC, Canada, V6T 1Z4
[b]Department of Integrative Oncology, BC Cancer Research Institute, BC, Canada, V5Z 1L3
[c]Department of Radiology, University of British Columbia, BC, Canada, V5Z 1M9

*Abstract* — **Background and Objective:** Training computer-vision related algorithms on medical images for disease diagnosis or image segmentation is difficult in large part due to privacy concerns. For this reason, generative image models are highly sought after to facilitate data sharing. However, 3-D generative models are understudied, and investigation of their privacy leakage is needed. **Methods:** We introduce our 3-D generative model, Transversal GAN (TrGAN), using head & neck PET images which are conditioned on tumour masks as a case study. We define quantitative measures of image fidelity and utility, and propose a novel framework for evaluating privacy-utility trade-off through membership inference attack. These metrics are evaluated in the course of training to identify ideal fidelity, utility and privacy trade-offs and establish the relationships between these parameters. **Results:** We show that the discriminator of the TrGAN is vulnerable to attack, and that an attacker can identify which samples were used in training with almost perfect accuracy (AUC = 0.99). We also show that an attacker with access to only the generator cannot reliably classify whether a sample had been used for training (AUC = 0.51). We also propose and demonstrate a general decision procedure for any deep learning based generative model, which allows the user to quantify and evaluate the decision trade-off between downstream utility and privacy protection. **Conclusions:** TrGAN can generate 3-D medical images that retain important image features and statistical properties of the training data set, with minimal privacy loss as determined by a membership inference attack. Our utility-privacy decision procedure may be beneficial to researchers who wish to share data or lack a sufficient number of large labeled image datasets.

## I. INTRODUCTION

Medical imaging tasks, such as disease diagnosis, image segmentation, image reconstruction, and other related functions, are increasingly benefiting from automation and the integration of deep-learning models [1][2][3][4]. Training these models is data-intensive due to the substantial number of parameters that require learning and the size of the image datasets. Additionally, the need for annotated samples often necessitates expertise in a specific dataset. While data sharing among institutions is a potential solution to overcome data requirements, privacy concerns can make it challenging [5]. Availability of public medical datasets is often limited, and the quality of available datasets may vary. Moreover, the rarity of certain disease types results in imbalanced datasets.

One possible solution to these problems is federated learning[6][7]. For example, Sheller et al. [8] are among the first researchers who applied federated learning to solve a medical image segmentation problem. However, federated learning has limitations: 1) In some cases, it has been demonstrated that patient metadata can be reconstructed by examining the model's outputs and structure trained in a federated setting [9]. 2) Legal restrictions in some jurisdictions may prohibit the use of medical images for cross-institutional model training, even with complete anonymization, patient consent, and institutional approval. 3) A separate legal agreement may be required for each entity interested in using imaging data from a particular institution. 4) Data from different institutions may exhibit high heterogeneity and may not follow an independent and identically distributed pattern [10][11]. An alternative solution is based on synthetic image generation. Since many studies often analyze data at the cohort level rather than for individual patients, synthetic images with similar properties (feature distributions and co-distributions) to real images can be utilized to share valuable information about the original dataset without compromising sensitive patient information.

Generative Adversarial Networks (GANs) represent one category of methods capable of generating realistic medical images [12][13]. GANs consist of a generator and a discriminator engaged in an adversarial game: the generator attempts to produce images that deceive the discriminator, while the discriminator aims to distinguish between generated/fake and real images. GANs have shown promise in several medical imaging studies. For example, they have been used to generate synthetic abnormal MRI images with brain tumors, synthesize high-resolution retinal fundus images, and generate synthetic pelvic CT images [14][15][16]. Within the PET imaging domain, GANs have also been employed to generate synthetic 2-D brain images [17].

It is well recognized that GAN-based models are susceptible to membership inference attacks [18], which seek to determine whether one or more test samples were part of the training data. These attacks exploit the fact that models often overfit to

*Corresponding author e-mail:* Robert.bergen@ubc.ca.

training data, allowing attackers to probe the model's discriminator or generator on test samples. Confirming training set membership can lead to the inference of sensitive information, such as disease diagnosis. Therefore, it is crucial to evaluate how membership attacks may compromise private information in generative models. Typically, attack performance is assessed using receiver operating characteristic (ROC) curves and area under the curve (AUC) metrics, although it is advisable to employ a metric tailored to the specific use case.

GANs and other machine learning models are susceptible to membership inference attacks due to their tendency to overfit to training data. Ensuring privacy is challenging and can come at the expense of model utility. Recently, Wang et al. introduced a method to provide a minimum privacy guarantee by regenerating data until it meets a predefined minimum utility and privacy threshold [20]. Xing et al. proposed a method for privacy quantification on synthetic X-ray images, dependent on detecting copies of real samples in the synthetic dataset [21]. Nevertheless, privacy breaches can still occur in the absence of copied real samples.

We propose a novel framework to evaluate a GAN's privacy leakage through two membership attacks. One attack simulates an attack on the discriminator, while the other simulates an attack on the generator. To demonstrate the privacy evaluation, we select a recently introduced GAN architecture capable of generating 3-D head & neck PET images [22]. This method has been shown to generate 3-D images with tumors defined by an input mask and has been evaluated in various ways, including radiomic analyses and segmentation training tasks. This is convenient because segmentation performance metrics offer one way to quantify a model's fidelity and utility. We visualize the relationships between the GAN's fidelity, utility, and privacy protection as a function of the number of training iterations.

Finally, we propose a decision procedure to evaluate the utility-privacy trade-off for any GAN. Since our method of quantifying privacy protection is not dependent on architecture, data type, or down-stream task, it can be easily generalized.

## II. METHODS

### A. Transversal GAN

The simplest GAN architecture consists of a generator and a discriminator playing an adversarial game (Figure 1). The generator synthesizes fake data while the discriminator tries to determine between real and fake data. The output classifications are used to update the generator on the next iteration so that it learns to generate higher quality data over time.

The GAN architecture used in this work is a modified version of the temporal GAN (TGAN) [23], originally developed for video generation. The TGAN breaks the generator into two components; a temporal generator, $G_0$, and an image generator, $G_1$ (Figure 2). The temporal generator takes a random latent variable $z_0$ as input and generates a temporal latent vector $z_1(t)$. The image generator generates frames of a video at time $t$ using the $z_0, z_1(t)$ variables as input. That is, a $T$-frame video is represented as the time series $[G_1(z_0, z_1(t = 1)), ..., G_1(z_0, z_1(t = T))]$. To improve stability of the training process, the spectral norm of each weight parameter in each layer is constrained to less than 1, which Saito et al. refer to as singular value clipping[23].

In our experiment, we apply a modified version of the TGAN on 3-D head and neck PET images, substituting the time dimension in videos for the 3$^{rd}$ spatial (axial) dimension in the 3-D volumes. The images are conditioned on masks, allowing us to generate 3-D volumes with specific tumour geometry. The TGAN has a hyperparameter, $\omega$, which determines how strongly the conditional tumour mask information is weighted relative to the input image in the discriminator.

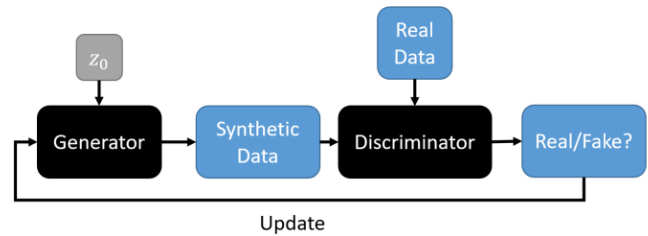

**Fig. 1.** *A simple GAN consists of a generator, synthesizing fake data from input noise, and a discriminator, which distinguishes between real and fake data. The generator and discriminator are updated iteratively, producing more realistic synthetic data over time.*

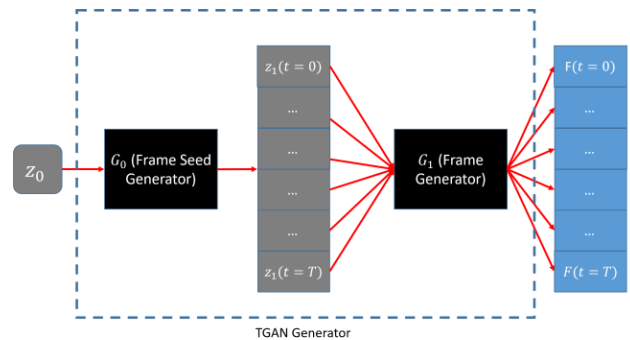

**Fig. 2.** The TGAN generator for video generation is broken up into two components. $G_0$ learns the temporal dynamics of the video to generate frame seeds while $G_1$ is an image generator that generates individual frames.

The weighting of image $I$ and mask $M$ is given by $I' = (1 - \omega)I$, $M' = \omega M$. We use a value of $\omega = 0.01$ as suggested by Bergen et al [14]. The model was trained for 5000 epochs (23 hours) using the RMSProp optimizer with learning rate 0.00005, and Wasserstein loss on a P40 GPU (24 GB) with a batch size of 32. We perform singular value clipping every 5 iterations. Since the transversal direction is used as a proxy for the temporal component in TGAN, we name our modified version the Transversal GAN. Our code for training the GAN can be found at https://www.github.com/robbergen/TrGAN.

### B. Segmentations

To segment our real and synthetic tumours, we use the same neural network segmentation architecture as the winner of the MICCAI 2020 Head and Neck Tumor (HECKTOR) segmentation challenge, as described by Iantsen et al [24]. This network is designed on the U-net architecture with residual layers and supplemented with squeeze-and-excitation normalization. The only modification we make is the number of inputs. The original architecture takes in two inputs; the head & neck PET image as well as a CT image. Our version only accepts PET images as input. The models were trained for 150 epochs using a P40 GPU (24 GB) with a batch size of 2. Segmentation quality was evaluated using the DICE score (DSC) metric.

### C. Privacy Assessment

To assess the privacy leakage in the Transversal GAN, two different membership inference attacks were conducted. The first attack assumes that the attacker has access to the model's discriminator. In this scenario, the assumptions made are:

- The attacker has access to $n = 200$ images, of which $m = 150$ were used for training the Transversal GAN.
- The attacker has access to the discriminator.

For classification, the attacker calculates the discriminator output for each sample and ranks them accordingly. Since the discriminator uses a Wasserstein loss function, the output is a real number $\in [-1,1]$. If the discriminator overfits, it will tend to score training samples with higher scores than non-training samples, and an inference on training set membership can be made.

In the second attack scenario, it is assumed that the attacker uses the generator to make the membership inference attack instead of the discriminator. Liu et al describe a membership attack schema on generative image models which is adopted here [25]. To summarize, the attacker trains a network, A, to learn latent vectors $z_A$ for the generator. The attack network takes in an image sample $x_i$ as input. The generated image and image sample $I$ are compared and the L2 norm for sample $i$ ($L_i$) is used as the loss metric for the attack network.

$$L_i = \left\| G\big(A(x_i)\big) - x_i \right\|_2 \qquad (1)$$

The attack network consists of 2 dense layers with relu activation. It is trained for 10000 iterations and the minimum loss value $L_{min}(x_i)$ over all iterations is calculated for each sample. This process is repeated four times to reduce initialization effects. The rationale here is that if the generator overfits, the attack network is able to learn latent vectors that closely reproduce training samples and the reconstruction loss will be low. Similar to the discriminator attack scenario, the minimum loss values are used to rank and classify samples as training or non-training samples. It is important to note that the attacker has full access to the generator internals, making it possible to compute the gradient of the reconstruction distance w.r.t. the attack network parameters analytically.

To quantify privacy leakage, the discriminator and generator attack outputs were calculated for the true training and non-training distributions. Then, a t-test was performed to determine whether the two distributions had any statistically significant differences. Then, receiver operating characteristic (ROC) curves were generated to evaluate the performance of the attacker classifications.

### D. Fidelity

Synthetic data fidelity is measured to determine how faithfully the synthetic data distribution represents the original data distribution. To measure fidelity, eight radiomic features are calculated over segmented tumours in both real and synthetic data distributions using the Py-Radiomics software package[26]. These features are Metabolic Tumour Volume (MTV), maximum, mean and peak measurements of the standardized uptake values (SUV), total lesion glycolysis (TLG) and 3 metrics based on the grey level co-occurrence matrix (GLCM), namely energy, entropy and homogeneity. Correlations are calculated between feature pairs within the real populations (N=201) and synthetic populations (N=201). The synthetic and real correlation coefficients are compared in two ways. The motivation behind this is to ensure that strong feature correlations that exist in the real data are preserved in the synthetic data. The correlations are compared in two ways. First, the mean squared error (MSE) between the coefficients is calculated, and second, by calculating a correlation accuracy score [27]. This is done by binning the correlation coefficients, into 5 equal bins and calculating how accurately the synthetic coefficients are binned using the real binned data as a gold standard. The correlation accuracy and MSE are plotted as a function of training iterations to visualize how fidelity changes over time.

### E. Privacy vs Utility

While fidelity measures how faithfully synthetic data reproduce a given real distribution of data, they do not assess how useful the data are for specific tasks. Utility can be quantified by using the synthetic data for a specific task and comparing the results with an identical experiment carried out with real data.

It is of interest to visualize how a model's privacy protection and utility change over the course of training. Ideally, a model has both high privacy protection and has high utility. However, high utility models often overfit on training data and are therefore not private. A model with high privacy protection could be constructed by introducing a large amount of regularization during training, but this ultimately reduces the model's utility. By plotting the model's utility and privacy protection as a function of training iteration, we can gain a better understanding of the utility-privacy protection trade-off.

To quantify utility, the Transversal GAN is used to generate images $I_{syn}$ from real tumour masks $M$. Then, a segmentation model, $S_{syn}$ is trained using the synthetic images. A validation set of real images and masks are then tested by calculating the average DSC between the masks and the segmentation output.

$$U = DSC(M_{real}, S_{syn}(I_{real})) \qquad (2)$$

The measure of utility defined above reflects the ability of the Transversal GAN model to train a segmentation model. A good value of utility is one which approaches or exceeds the utility of the same segmentation model trained on real data.

An alternate method is to measure relative utility increase by measuring the increase in DSC after synthetic data has been added to a real data set. Here, $S_{real}$ is a segmentation model trained on real images and evaluated on a test set of images $I_{real}$. $S_{synandreal}$ is trained on both synthetic and real images and evaluated on $I_{real}$.

$$U = DSC\left(M_{real}, S_{synandreal}(I_{real})\right) - \qquad (3)$$
$$DSC(M_{real}, S_{real}(I_{real}))$$

We measure this quantity when the training set for synthetic data generation consists of 135 images and the training set for $S_{real}$ consists of 40 images, simulating a scenario where a researcher has only a small amount of real data available to them, but has access to a larger database of synthetic images. 26 images were used as a test set for the segmentation models.

Privacy protection is quantified relative to the AUC metric for the membership attacks. For a given attack, the privacy protection is defined as

$$P = 2 * (1 - AUC) \qquad (4)$$

### F. Data

We utilized a publicly available dataset in The Cancer Imaging Archive (TCIA), further refined within the MICCAI 2020 Head & Neck Tumor (HECKTOR) challenge [28]; it comprises 201 cases from four centers. Each case comprises a PET image and GTVt (primary Gross Tumor Volume) mask, as well as a bounding box location. We use the bounding box information to crop the PET and GTVt masks to 64x64x32 volumes for input into the Transversal GAN and segmentation networks. The in-plane (axial) resolution of the PET images ranged from 3.5 mm to 3.9 mm while the through-plane resolution was 3.7mm. After cropping, this corresponds to a minimum field of view of (224 mm × 224 mm × 118.4 mm). In the pre-processing stage, images were normalized to values between [-1,1]. For assessing privacy leakage in the Transversal GAN, 150 images were used for training the GAN, and 50 were randomly withheld for evaluating the utility (Eq. 2) and privacy (Eq. 4).

## III. RESULTS

### A. Synthetic Data

The PET images generated by the conditional Transversal GAN are volumes of size 64x64x32. We observed that the shape of the head and internal anatomy structure all appeared realistic. The conditional Transversal GAN reproduces lesions based on the input tumour mask (Figures 3,4). Figure 3 shows images generated by masks that were seen by the generator during training, while Figure 4 shows images generated by masks that were not seen during training. It appears that there is good generalization and there are no visibly noticeable issues when generating images given unseen tumour masks.

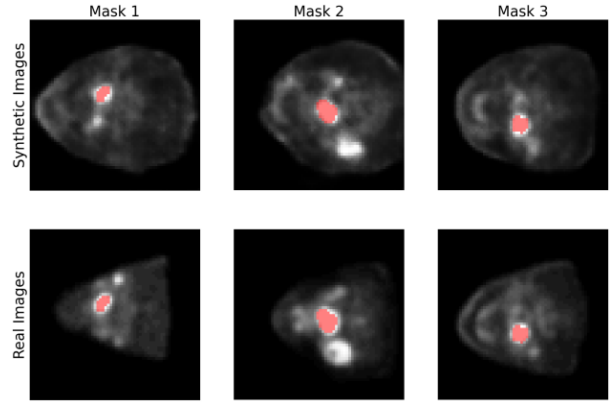

**Fig. 3.** Examples of real and synthetic PET images with tumour masks overlaid in red. Each column shows the synthetic image generated by a tumour mask and the corresponding real PET image. The tumour masks shown here were included in the training set.

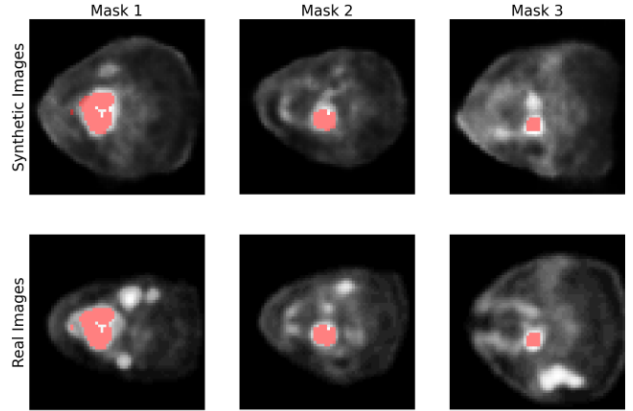

**Fig. 4.** Examples of real and synthetic PET images with tumour masks overlaid in red. Each column shows the synthetic image generated by a tumour mask and the corresponding real PET image. The tumour masks shown here were not included during training.

Theoretically, the Transversal GAN could be used to generate higher-resolution volumes and our initial tests show that training is also stable at 64x64x64 volumes. We chose to crop our images at 64x64x32 because of memory constraints for our system and long training times for larger volumes. Despite the small volume size, we are still able to encapsulate the entire head and neck for each patient with this crop setting.

### B. Segmentations

We generated 200 synthetic volumes on the conditional Transversal GAN, then performed automatic segmentations on them. The Dice scores for synthetic images were calculated and compared to the user-defined masks. The same segmentation algorithm was used on the real data for comparison. The average Dice score for real and synthetic data was 0.7 and 0.65, respectively. As a reference point, Iantsen et al achieved a 0.759 Dice score when using additional CT data for guidance [24]. Synthetic Dice score distributions were similar to real Dice score distributions. Figure 5 shows some examples of real and synthetic tumour segmentations, along with the original masks used to synthesize the images. Real and synthetic data are displayed at identical window and level settings.

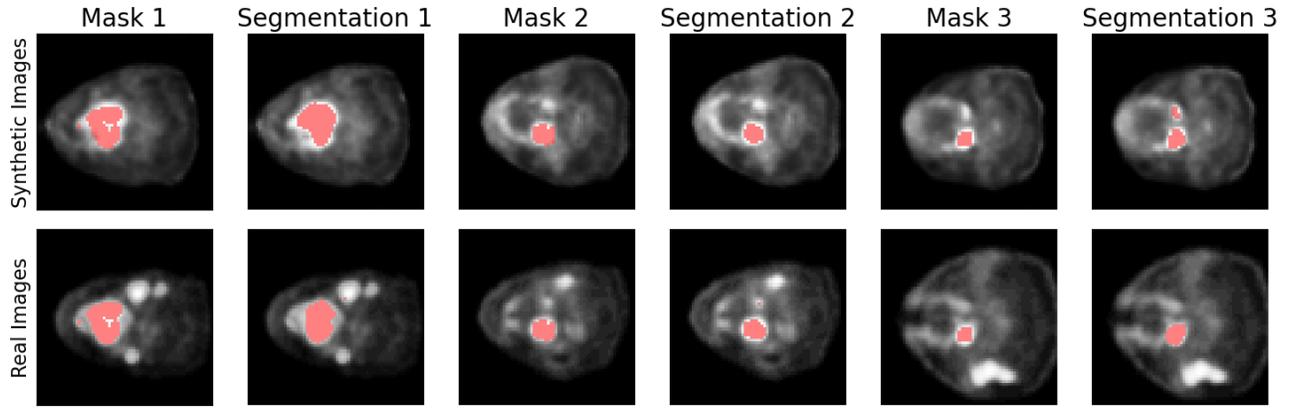

**Fig. 5.** Fidelity evaluation. Columns 1, 3, 5: Randomly selected examples of real and synthetic PET images with tumour masks overlaid in red. These columns show the synthetic image generated by a tumour mask and the corresponding real PET image. Columns 2, 4, 6: Examples of tumour segmentations for real and synthetic PET images.

### C. Fidelity

Changes in image fidelity are visualized in Figure 6. Correlation accuracy reaches its maximum at the same point that correlation MSE reaches its minimum, at 18000 iterations. Although fidelity appears to drop slightly after 18000 iterations, there were no visible drops in quality in the synthetic images as training continued.

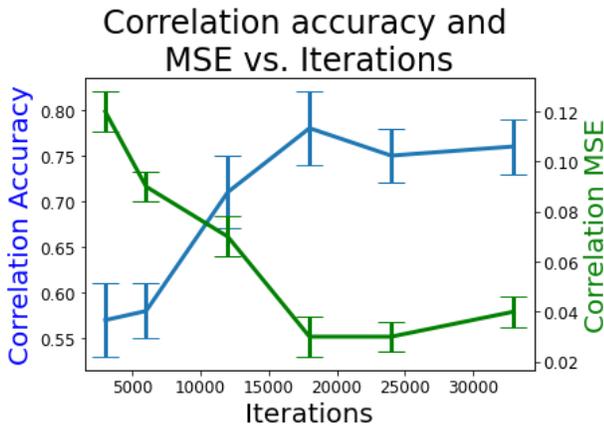

**Fig. 6.** Correlation accuracy (left y-axis; blue) and correlation MSE (right y-axis; green) during training. Results are averaged over 5 runs with errorbars representing the standard deviation of correlation accuracy and MSE.

### D. Privacy Assessment: Discriminator Attack

Histograms of the discriminator outputs for all training and non-training samples are shown in Figure 7. The discriminator output is bounded by $[-1,1]$ and assigns higher values to images in the training set. There is a statistically significant difference in the two distributions ($p < 0.0001$). The distributions of the training and non-training samples are so distinct that almost every sample can be classified with extremely high confidence. This indicates that the discriminator is very vulnerable to membership inference attacks. The ROC curve for the attack classifier is shown in Figure 8. The AUC was 0.999 indicating heavy privacy leakage.

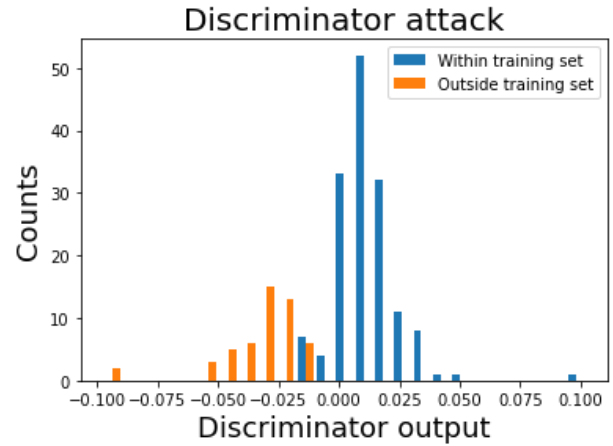

**Fig. 7.** Histogram of discriminator outputs for the discriminator membership inference attack. There is a clear distinction between training and non-training sample distributions, indicating privacy leakage.

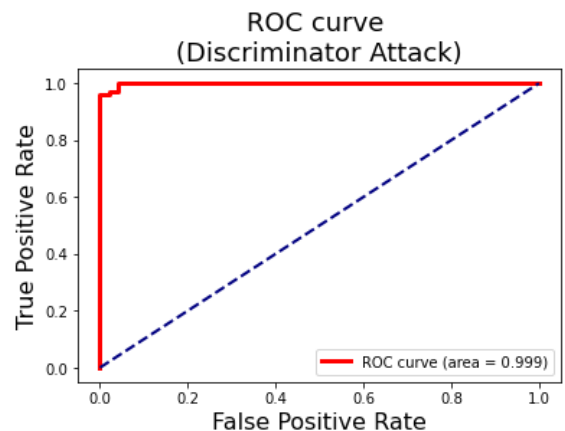

**Fig. 8.** ROC curve for the discriminator attack classifier after 33000 iterations.

### E. Privacy Assessment: Generator Attack

The attack network was used to learn latent vectors in order to reconstruct image samples using the generator. The minimum loss values of the generator attack network were calculated for each sample after training the Transversal GAN for 33000 iterations, and the histogram of these values is plotted in Figure 9. Unlike the discriminator, the training and non-training distributions are not significantly different ($p = 0.37$). As seen in Figure 10, the ROC curve for the classifier in this case performs approximately as well as random guessing, with an AUC of 0.51, indicating that the privacy leakage was minimal. In fact, even when the attacker knows the number of training samples ($m = 150$) in the distribution, the classifier accuracy was found to be 0.78, which is not significantly different from a dummy model that predicts all samples as training samples (accuracy = 0.75).

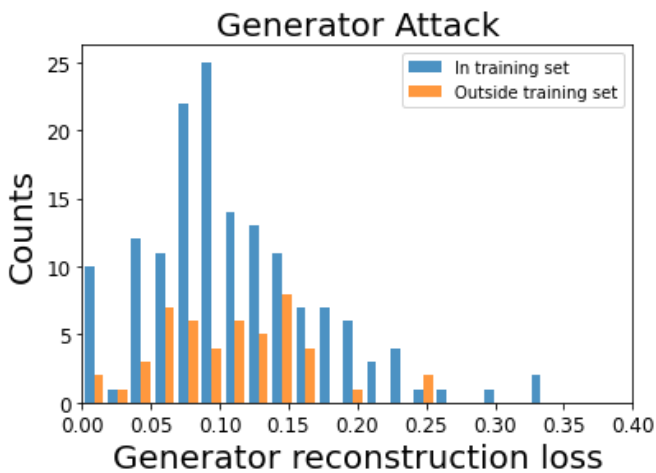

**Fig. 9.** Histogram of minimum losses in the generator attack. There is no statistically significant differences between training sample and non-training sample distributions.

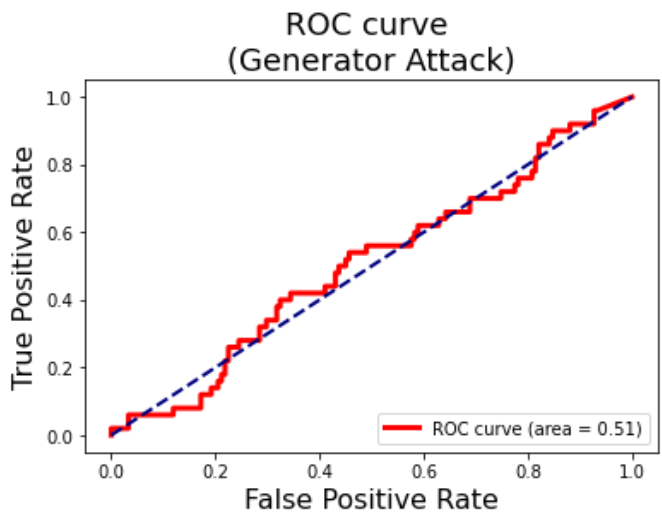

**Fig. 10.** ROC curve for the generator attack classifier after 33000 iterations.

### F. Privacy vs Utility

The discriminator privacy and utility (Equations 2, 4) for the model was calculated using during training and is shown in Figure 11. Utility slowly increases as expected, as images become more realistic. Privacy protection falls to 0 as the discriminator overfits and the AUC of the discriminator attack approaches 1. The ideal privacy and utility tradeoff point appears to be at around 12000 iterations, which corresponds to the elbow of the privacy-utility curve in Figure 12. The dashed blue line shows the privacy protection of the generator using the MIA on the generator described in Section 2C. Using the alternate measure of relative utility increase in Eq. 3., we found the increase in utility was U = 0.06. This represents an increase in DSC from 0.58 to 0.64 after synthetic data was introduced into the segmentation model training.

At this point a distinction between fidelity, utility and 'realism' must be made. As the Transversal GAN is trained, it produces more realistic images by learning the distributions of both the lesions and healthy anatomy. The utility is defined strictly in terms of the lesion segmentation, and it is unclear how the 'realism' of the surrounding healthy anatomy impacts this metric. Figures 9 and 10 imply an ideal privacy/utility trade-off point at approximately iteration 15000, but they do not imply that there is an equivalent trade-off of realism and privacy.

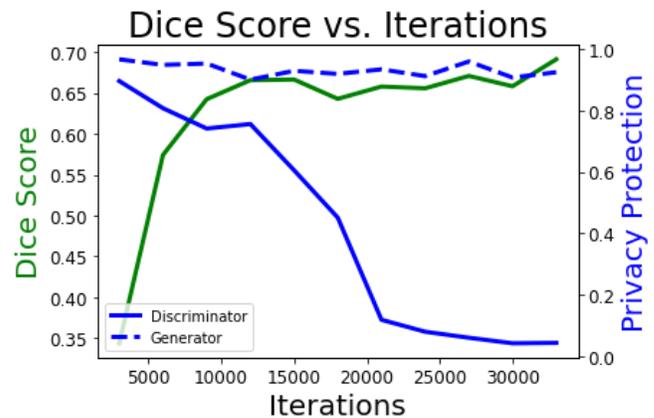

**Fig. 11.** Privacy and utility of the Transversal GAN model vs training epochs.

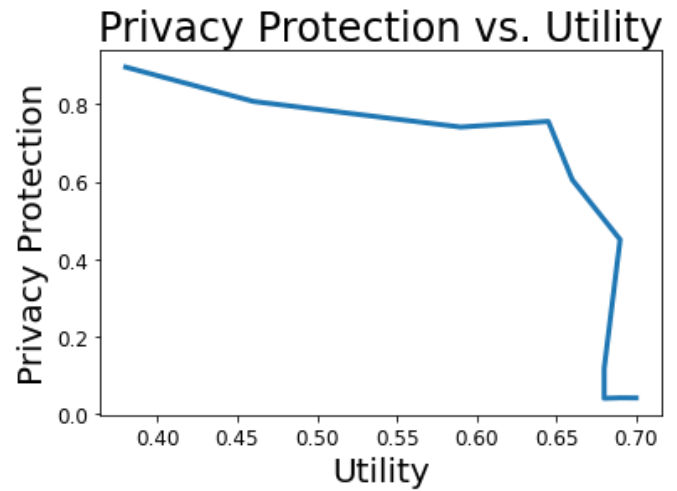

**Fig. 12.** Privacy-utility curve for the Transversal GAN discriminator.

Figures 13 and 14 illustrate the difference between synthetic images generated from the Transversal GAN model at iterations 15000 and 30000. From a utility perspective, the models are almost identical. On the other hand, the image at 30000 iterations is, at least upon visual inspection, more realistic. These examples highlight the fact that utility is highly task-dependent. A researcher wishing to use a synthetic Transversal GAN dataset for another downstream task, such as training a disease classifier, cannot rely on the same utility curve for choosing the optimal privacy-utility trade off.

There is no consensus on a universal definition of utility for medical images. Metrics like Inception Score or Frechet Inception Distance rely on pre-trained networks which have not incorporated medical images into the training process. These metrics may therefore be misleading when evaluating medical image realism. The best method for quantifying the overall realism of a synthetic data set remains an open question.

One possible approach to evaluate the realism of simulated images is using a two-alternative forced-choice (2AFC) detection paradigm with human observers. Within the forced-choice detection paradigm, the observer is presented with paired images (e.g., one real and one synthetic) and is asked to differentiate between them. In a related work, Fedrigo et al. [29] recently performed an 2AFC observer study on our TrGAN images, in which real/synthetic image pairs were presented as 2D-transaxial slices. Six out of eight observers could not identify the real image with statistical significance, indicating that the synthetic dataset was reasonably representative of oncological PET images. In future studies, we will evaluate image realism for different observers (e.g., technologists, senior physicists) and study paradigms, as well as consider volumetric images for more convenient evaluation.

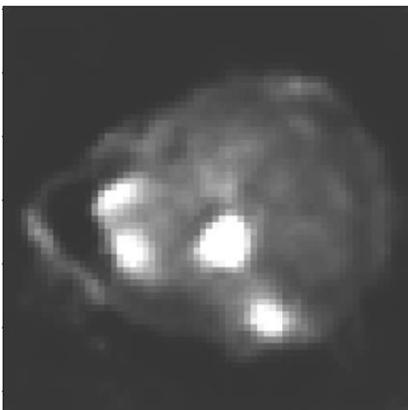

**Fig. 13.** Synthetic image generated by Transversal GAN after 15000 iterations. The utility of the model was measured as 0.65, while its privacy protection was measured as 0.6.

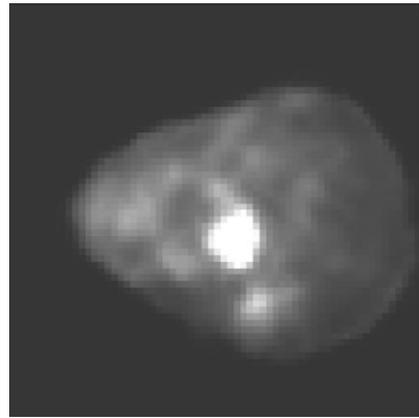

**Fig. 14.** Synthetic image generated by Transversal GAN after 30000 iterations. The utility of the model was measured as 0.68, while its privacy protection was measured as 0.05.

### G. A decision procedure for evaluating the optimal utility-privacy trade-off

Publicly releasing generative models trained on health data is potentially complicated because with higher utility can lead to a greater risk of privacy leakage. To assist researchers in making these trade-offs, we propose a general decision procedure to determine the optimal trade-off. In this discussion, we evaluate this trade-off for a model trained for varying number of epochs, but this procedure is also valid for any hyper-parameter (or combination of hyper-parameters) of the model.

Our proposed measure of privacy protection is defined by the performance of a membership inference attack, which can be carried out for any model architecture or downstream task. We use the fact that an AUC of 0.7 or below is generally considered poor [30], to define the minimum acceptable privacy protection threshold for any model as 0.6 (Eq. 4).

A minimum threshold for model utility is not possible to generalize, since utility is highly dependent on the model's down-stream task. It is also dependent on the context of the down-stream task (minimally acceptable metrics for image segmentation on medical images could vary substantially depending on the anatomical location, for example). However, we make the assumption that, given a down-stream task and its context, some minimum threshold can be meaningfully defined. Therefore, for the following illustrative example, we arbitrarily set the minimum threshold for utility as 0.6.

Given a set of points $\mathbf{X}$ on the utility-privacy protection plane, measured at different number of training iterations, we can now define a set of candidate points $\mathbf{C}$ as

$$C = Conv(X_{thresh}) \tag{5}$$

where $X_{thresh} \subseteq \mathbf{X}$ is the subset of $\mathbf{X}$ that lies within the thresholded region of the utility-privacy protection plane and Conv($\mathbf{X_{thresh}}$) is the convex set of $\mathbf{X_{thresh}}$. Figure 15 shows an example where a set of candidate points are chosen. If there are several candidate points, we recommend choosing either the point with maximal utility, or the point that maximizes utility plus privacy.

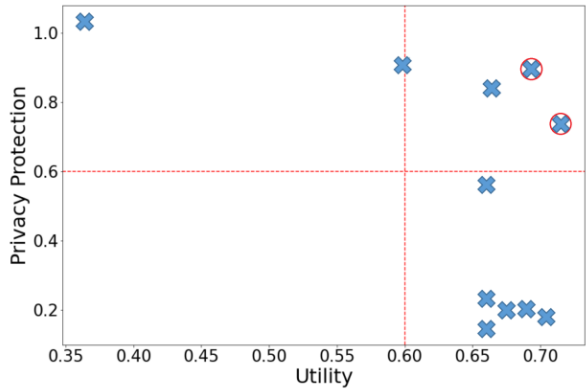

**Fig. 15** An example of a set of points on the utility-privacy protection plane. The two candidate points calculated by Eq. 5 are shown circled. Either point could be considered optimal according to Eq. 6.

$$x_{optimal} = argmax_{x_i}(U(x_i)) \quad \text{or}$$
$$x_{optimal} = argmax_{x_i}(U(x_i) + P(x_i)) \quad (6)$$

### H. Discussion

Previous quantitative analysis has shown synthetic data generated by TrGAN exhibits similar segmentation performance (based on Dice score) to real data, preserving strong statistical correlations between radiomic features observed in real data sets [22]. An advantage over traditional 3-D generative models is its ability to partially reduce memory consumption by decomposing the 3-D generator into a 2-D and 1-D generator. This study shows that this model's generator is potentially resilient to membership inference attacks while maintaining good image fidelity and utility. This is a significant finding, since generators can sometimes be vulnerable in the small training set size regime, in which this study was carried out [19]. The TrGAN model, therefore, holds promise for facilitating data sharing with minimal privacy risks. However, it is important to note that MIAs provide empirical evaluations of privacy but not guarantees. The results presented here represent an upper bound on the privacy of the GAN. It is possible that other attacks on the generator could be more successful, and further studies should be conducted.

The task-based definition of utility utilized in this study focused on segmentation; however, there is considerable potential to adapt this synthetic data for other applications, such as PET disease classification. For example, two active areas of research are distinguishing between Alzheimer's and Mild Cognitive Impairment (MCI)[31][32]. Another potential application of Transversal GAN is reducing the imbalance of data sets that contain underrepresented diseases or classes.

It has been well established that discriminators are the most vulnerable component of the GAN model[33]. Our model is no exception. Privacy would be a major concern if the discriminator was made available to an attacker. Although we conducted a discriminator attack analysis for completeness, it is essential to note that this is not a very realistic scenario. Most generative models will be used in either one of two ways: either the model generates synthetic data released to the public, or the generator is released, and other institutions may generate data as needed. Releasing the discriminator model yields no real benefit regarding data sharing or privacy.

Our results demonstrate that releasing the generator publicly entails very little privacy risk. One potential limitation in our analysis is that the number of samples available for training and testing was small (n = 201). An increase in sample size is likely to reduce privacy leakage in the discriminator, as overfitting would be diminished. Increased sample sizes have been shown to reduce privacy leakage in the generator as well [18]. Our promising results at low sample sizes suggest that our generator may be very robust against membership inference attacks, but further studies should be conducted to confirm this.

Another limitation of our attack model is that computing the reconstruction loss in Eq 1 is a non-convex problem. To reduce initialization effects we followed the guidelines of Liu et al [25] and ran the experiment multiple times, taking the lowest loss values. It is possible that with many more repetitions the accuracy of the attack network could improve somewhat. It is also unclear whether a deeper attack network architecture would improve the generator attack performance.

Recently it has been suggested that the ROC curve and AUC metrics are not sufficient to characterize privacy of a generative model. Carlini et al[34] suggest that performance of membership inference attacks should be heavily weighted at low false positive rates and recommend plotting the ROC curve on a logarithmic scale. We note that our generator performs well across all values of the false positive rate. Using this modified metric, our generator can still be considered robust against membership inference attack.

We conducted membership inference attacks that rely on an attacker having access to either the discriminator or the generator. However, GANs have also been shown to be vulnerable to other 'black box' attacks requiring only knowledge of input/output pairs [19]. While this type of attack was not evaluated in this study, it is important to note that these attacks are generally more challenging and less accurate, since the attacker does not have access to the internal GAN structure for attack optimization purposes.

There are several planned improvements to the Transversal GAN architecture in the future. For example, further improvements have been made to the TGAN architecture, on which Transversal GAN was based, in a newer version, TGANv2[35]. One major improvement is the availability of multi-GPU capable code. Another major modification is made to the training and inference processes in TGANv2. The TGANv2 generator is able to output sparse samples for training at lower computational cost and dense samples for inference. Future work will test the capability of TGANv2 to generate higher resolution 3-D medical images.

Other non-GAN based video generation techniques may be suitable for 3-D image generation as well. For example, Yan et al. use vector quantized variational auto encoders to compress the video data and perform the generation in the compressed representation using transformers[36]. Further research is needed to assess how transformer-based approaches compare to our method for 3-D image generation and evaluate their impact on privacy.

## IV. CONCLUSION

GAN-based models have shown promise in the field of medical image generation, especially for image-to-image translation tasks. Generating stand-alone 3-D medical images, on the other hand, is a technically challenging and resource-intensive task that has not been well researched. TrGAN can generate 3-D medical images that retain important image features and statistical properties of the training data set. We have shown that the discriminator is very vulnerable to membership inference attack, while the generator showed minimal privacy loss. Our utility-privacy decision procedure may be beneficial to researchers who wish to share data with other institutions, and for researchers to train classification or segmentation algorithms, but lack a sufficient number of large labeled image datasets.

## ACKNOWLEDGMENT

The authors wish to acknowledge the Natural Sciences and Engineering Research Council of Canada (NSERC) Discovery Grant RGPIN-2019-06467, and the BC Cancer Foundation.